\documentstyle[aps,amsfonts,amssymb]{revtex}
\begin{document}
\draft
\title{Relativistic Gamow Vectors}
\author{A.~Bohm, H.~Kaldass and S.~Wickramasekara}
\address{Department of Physics, University of Texas at Austin}
\author{P.~Kielanowski}
\address{Departamento de F\'{\i}sica, Centro de Investigaci\'on
y de Estudios Avanzados del IPN, Mexico City}
\maketitle
\begin{abstract}
Gamow vectors in non-relativistic quantum mechanics are generalized
eigenvectors (kets) of self-adjoint Hamiltonians with complex
eigenvalues $(E_{R}\mp i\Gamma/2)$. Like the Dirac kets, they are
mathematically well defined in the Rigged Hilbert Space $\Phi\subset
{\cal H}\subset \Phi^{\times}$. Gamow kets are derived from the
resonance poles of the S-matrix $S_{j}(z)$ at $z=z_{R}=E_{R}\mp
i\Gamma/2$. They have a Breit-Wigner energy distribution, an
exponential decay law, and are members of a basis vector expansion
whose truncation gives the finite dimensional effective theories with
a complex Hamiltonian matrix. They also have an asymmetric time
evolution described by a {\it semigroup} generated by the Hamiltonian,
which expresses a fundamental quantum mechanical arrow of time. These
Gamow kets are generalized to relativistic Gamow vectors by
extrapolating from the Galilei group to the Poincar\'e group. This
leads to semigroup representations of the Poincar\'e group which are
characterized by spin $j$ and complex invariant mass square
${\mathsf{s}}={\mathsf{s}}_{R}=\left(
M_{R}-\frac{i}{2}\Gamma\right)^{2}$. In these non-unitary
representations $(j,{\mathsf{s}}_{R})$ the Lorentz subgroup is
unitarily represented and the four-momenta are ``minimally complex''
in the sense that the four-velocity
$\hat{p}_{\mu}=\frac{p_{\mu}}{\sqrt{{\mathsf{s}}_{R}}}$ is real.  The
relativistic Gamow vectors have all the properties listed above for
the non-relativistic Gamow vectors and are therefore ideally suited to
describe relativistic resonances and quasistable particles with
resonance mass $M_{R}$ and lifetime $\hbar/\Gamma$.
\end{abstract}
\section{Introduction}
Following Wigner~\cite{Wigner}, an elementary relativistic quantum
system, an elementary particle with mass $m$ and spin $j$ is in the
mathematical theory described by the space of an unitary irreducible
representation (UIR) of the Poincar\'e group~${\mathcal P}$. From
these UIR, the relativistic quantum fields are constructed
\cite{Weinberg}.  More complicated relativistic systems are described
by direct sums of UIR (for ``towers'' of elementary particles) or by
direct products of UIR (for combination of two or more elementary
particles) \cite{Weinberg}.  A direct product of UIR may be decomposed
into a continuous direct sum (integral) of irreducible representations
\cite{Joos,Macf}.  The UIR are characterized by three invariants
$(m^2, j,\textrm{sign}(p_0))$, where~$j$ represents the spin and the
real number $m$ represents the mass of elementary particle (we
restrict ourselves here to $\textrm{sign}(p_0)=+1$).

The UIR of the Poincar\'e group ${\mathcal P}$ describe stable
elementary particles (stationary systems). There is only a very small
number of truly stable particles in nature and most relativistic (and
also non-relativistic) quantum systems are decaying states (weakly or
electromagnetically), or hadron resonances with an exponential decay
law and finite lifetime $\tau_R=\frac{\hbar}{\Gamma}$ (in their rest
frame) and a Breit-Wigner energy (at rest) distribution. The UIR of
${\mathcal P}$ therefore describe only a few of the relativistic
quantum systems in nature; for the vast majority of elementary
particles listed in the Particle Physics Booklet~\cite{PPB}, the UIR
provide only a more or less approximate description.  We want to
present here a special class of (non-unitary) semi-group
representations of $\cal P$ which describe quasistable relativistic
particles.

Phenomenologically, one always takes the point of view that resonances
are autonomous quantum physical entities, and decaying particles are
no less fundamental than stable particles. Stable particles are not
qualitatively different from quasistable particles, but only
quantitatively by a zero (or very small) value of $\Gamma$. Therefore
both stable and quasistable states should be described on the same
footing. This has been accomplished in the non-relativistic case,
where a decaying state is described by a generalized eigenvector of
the (self adjoint, semi-bounded) Hamiltonian with a complex eigenvalue
$z_R=E_R-i\Gamma/2$~\cite{Bohm1} called Gamow vectors. The stable
state vectors with real eigenvalues $E_S$ are the special case with
$\Gamma=0$, i.e., $z_R=E_R-i\Gamma/2\rightarrow E_S$.

In the standard Hilbert space formulation of quantum mechanics, such
vectors do not exist and one had to employ the Rigged Hilbert Space
(RHS) formulation of quantum mechanics. Dirac's bras and kets are,
mathematically, generalized eigenvectors with real eigenvalues, and
Gamow vectors are generalizations of Dirac kets. They are described by
kets $\psi^G\equiv | z_R^-\rangle\sqrt{2\pi\Gamma}$ with complex
eigenvalue $z_R=E_R-i\Gamma/2$, where $E_R$ and $\Gamma$ are
respectively interpreted as resonance energy and width. Like Dirac
kets, the Gamow kets are functionals of a Rigged Hilbert Space :
\begin{equation}
\Phi_+\subset{\mathcal H}\subset\Phi^\times_+:
\,\,\,\,\,\,
\psi^G\in\Phi^\times_+,
\label{eq1}
\end{equation}
and the mathematical meaning of the eigenvalue equation
$H^\times|z_R^-\rangle=(E_R-i\Gamma/2)|z_R^-\rangle$ is:
\begin{equation}
\langle H\psi|z_R^-\rangle
\equiv
\langle\psi|H^\times|z_R^-\rangle=
z_R\langle\psi|z_R^-\rangle
\,\,\,\,\,
\textup{for all}
\,\,\,\,\,
\psi\in\Phi_+.
\label{eq2}
\end{equation}
The conjugate operator $H^\times$ of the Hamiltonian $H$ is uniquely
defined by the first equality in~(\ref{eq2}), as the extension of the
Hilbert space adjoint operator $H^\dagger$ to the space of functionals
$\Phi^\times_+$~\footnote{For (essentially) self adjoint $H$,
$H^\dagger$ is equal to (the closure of) $H$; but we shall use the
definition~(\ref{eq2}) also for unitary operators ${\mathcal U}$ where
${\mathcal U}^\times$ is the extension of ${\mathcal U}^\dagger$, but
not of ${\mathcal U}$.}; on the space ${\mathcal H}$, the operators
$H^\times$ and $H^\dagger$ are the same.

The non-relativistic Gamow vectors have the following properties:
\begin{enumerate}
\item
They have an asymmetric (i.e., $t\geq 0$ only) 
time evolution and obey then an exponential law:
\begin{equation}
\psi^G(t)=\textrm{e}_+^{-iH^\times t}
|E_R-i\Gamma/2^-\rangle=
\textrm{e}^{-iE_R t}
\textrm{e}^{-\Gamma t/2}
|E_R-i\Gamma/2^-\rangle,
\,\,\,\,
\textup{only for}
\,\,\,\,
t\geq0.
\label{eq3}
\end{equation}
There is another Gamow vector
$\tilde\psi^G=|E_R+i\Gamma/2^+\rangle\in\Phi^\times_-$, and another
semigroup $\textrm{e}_-^{-iH^\times t}$ for $t\leq0$ in another RHS
$\Phi_-\subset{\mathcal H}\subset\Phi_-^\times$ (with the same
${\mathcal H}$) with the asymmetric evolution
\begin{equation}
\tilde\psi^G(t)=\textrm{e}_-^{-iH^\times t}
|E_R+i\Gamma/2^+\rangle=
\textrm{e}^{-iE_R t}
\textrm{e}^{\Gamma t/2}
|E_R+i\Gamma/2^+\rangle,
\,\,\,\,
\textup{only for}
\,\,\,\,
t\leq0.
\label{eq4}
\end{equation}
\item
The $\psi^G$ ($\tilde\psi^G$) is derived as a functional at the
resonance pole term located at $z_R=(E_R-i\Gamma/2)$ (at
$z^*_R=(E_R+i\Gamma/2)$) in the second sheet of the analytically
continued S-matrix.
\item
The Gamow vectors have a Breit-Wigner energy distribution
\begin{equation}
\langle^-E|\psi^G\rangle=
i\sqrt{\frac{\Gamma}{2\pi}}\frac{1}{E-(E_R-i\Gamma/2)},
\,\,\,\,
-\infty_{II}<E<\infty,
\label{eq5}
\end{equation}
where $-\infty_{II}$ means that it extends to $-\infty$ on the second
sheet of the S-matrix (whereas the standard Breit-Wigner extends to
the threshold $E=0$).
\end{enumerate}
We want to present here a generalization of these non-relativistic
Gamow vectors to the relativistic case. 

In the non-relativistic case the inclusion of the degeneracy quantum
numbers of energy, i.e., the extension of the Dirac-Lippmann-Schwinger
kets
\begin{eqnarray}
&&|E^\pm\rangle=|E\rangle+\frac{1}{E-H\pm i0}V|E\rangle
=\Omega^\pm|E\rangle\nonumber\\[6pt]
&&H|E^\pm\rangle=
E|E^\pm\rangle;
\,\,\,\,
(H-V)|E\rangle=E|E\rangle
\label{eq6}
\end{eqnarray}
to the basis of the whole Galilei group is trivial.

For the two particle scattering states (direct product of two
irreducible representations of the Galilei group~\cite{ref5}) one uses
eigenvectors of angular momentum $(jj_3)$ for the relative motion and
total momentum ${\bbox p}$ for the center of mass motion. Thus
\begin{equation}
|E^{\textrm{\scriptsize tot}}{\bbox p}jj_3(l,s)
\,\,^\pm\rangle=
|{\bbox p}\rangle\otimes|Ejj_3\,\,^\pm\rangle
\label{eq7}
\end{equation}
where $E^{\textrm{\scriptsize tot}}=\frac{{\bbox p}^2}{2m}+E$ (the
Hamiltonian in~(\ref{eq6}) is $H=H^{\textrm{\scriptsize
tot}}-\frac{{\bbox P}^2}{2m}$).
The center-of-mass motion is usually separated by transforming to
the center-of-mass frame, and there one uses in~(\ref{eq6})
\begin{equation}
|{\bbox p}=\bbox{0}\rangle\otimes|Ejj_3\,\,^\pm\rangle
=|E,\bbox{p}=0,jj_3\,\,^\pm\rangle=
|E\,\,^\pm\rangle.
\label{eq8}
\end{equation}

The generalized eigenvectors
\begin{equation}
|Ejj_3\,\,^\mp\rangle\in\Phi^\times_\pm\supset{\mathcal H}\supset\Phi_\pm
\label{eq9}
\end{equation}
with
\begin{equation}
H^\times|Ejj_3\,\,^\pm\rangle=
E|Ejj_3\,\,^\pm\rangle,\quad 0\leq E <\infty,
\label{eq10}
\end{equation}
where $E$ runs along the cut on the positive real axis of the 1-st
sheet of the $j$-th partial S-matrix, are the scattering states. The
proper eigenvectors of~(\ref{eq9}) with $E=-|E_n|$ at the poles on the
negative real axis on the 1-st sheet are the bound states
$|E_njj_3\rangle\in\Phi$. By the Galilei transformation one can
transform these vectors to arbitrary momentum ${\bbox p}$; $E$ and
${\bbox p}$ are not intermingled by Galilei transformations.

To obtain the non-relativistic Gamow kets one analytically continues
the Dirac-Lippmann-Schwinger ket~(\ref{eq6}) into the second sheet of
the $j$-th partial S-matrix to the position of the resonance pole
$|z_R=E_R-i\Gamma/2,jj_3\ ^-\rangle$ and obtains the following
representation~\cite{Bohm1}:
\begin{equation}
|z_R=E_R-i\Gamma/2,jj_3\ ^-\rangle=
\frac{i}{2\pi}\int_{-\infty_{II}}^{+\infty}dE
|Ejj_3\ ^-\rangle\frac{1}{E-z_R}.
\end{equation}
A Galilei transformation can boost this Gamow ket to any \textit{real}
momentum~$\bbox{p}$
\[
|{\bbox p},z_R,jj_3\ ^-\rangle=
{\mathcal U}(\bbox{p})|\bbox{0}\rangle\otimes|z_Rjj_3\ ^-\rangle.
\]
However, complex momenta cannot be obtained in this way since the
Galilei transformations commute with the intrinsic energy
operator~$H$.

In the relativistic case the Lorentz transformation -- in particular
Lorentz boosts -- intermingle energy $E^{\textrm{\scriptsize
tot}}=p^0$ and momenta $p^m$, $m=1,2,3$.  Thus complex energy or
complex mass also leads to complex momenta. This has led in the past
to consider complicated complex momentum representations of the
Poincar\'e group ${\mathcal P}$. To restrict this unwieldy set of
Poincar\'e group representations we will consider a special class of
``minimally complex'' irreducible representations of ${\mathcal P}$ to
describe relativistic resonances and decaying elementary
particles. Our construction will also lead to complex momenta $p^\mu$,
but in our case the momenta will be ``minimally complex'' in such a
way that the 4-velocities $\hat{p}_\mu\equiv\frac{p_\mu}{m}$ remain
real. This construction was motivated by a remark of
D.~Zwanziger~\cite{zwanzi} and is based on the fact that the
4-velocity eigenvectors $|\hat{\bbox{p}}j_{3}(mj)\rangle$ furnish as
valid a basis for the representation space of ${\mathcal P}$ as the
usual Wigner basis of momentum eigenvectors
$|\bbox{p}j_{3}(m,j)\rangle$.  This means every state of an UIR
$(m,j)$, ($\phi\in\Phi\subset{\cal H}(m,j)\subset\Phi^{\times}$, where
$\Phi$ denotes the space of well-behaved vectors and $\Phi^{\times}$
the space of kets for the Hilbert space ${\cal H}(m,j)$ of an UIR),
can be written according to Dirac's basis vector decomposition as
\begin{equation}
\label{2.15a}
\phi=\sum_{j_{3}}\int \frac{d^{3}\hat{p}}{2\hat{p}^{0}}
|\hat{\bbox{p}},j_{3}\rangle\langle j_{3},\hat{\bbox{p}}|\phi\rangle
\end{equation}
where we have chosen the invariant measure
\begin{equation}
\label{measurehat}
d\mu(\hat{\bbox p}) = \frac{d^{3}\hat{p}}{2\hat{p}^{0}}
             = {\frac{1}{m^{2}}} \, {\frac{d^{3}p}{2 E({\bbox p})}},
\,\,\,\,\,\,     \hat{p}^{0} = \sqrt{1+\hat{\bbox p}^{2}} \, .
\end{equation}
As a consequence of (\ref{measurehat}), 
the $\delta$-function normalization of these velocity-basis vectors is
\begin{equation}
\langle \xi , \hat{\bbox p}\,|\,\hat{\bbox p}', \xi' \rangle
       = 2 \hat{p}^{0}  \delta^{3}(\hat{\bbox p}-\hat{\bbox p}')
                              \, \delta_{\xi \xi'}\\
 = 2 p^{0} m^{2} \delta^{3}
(\bbox{p}-\bbox{p'})\, 
\delta_{\xi \xi'} \, .
\label{normalizationhat} 
\end{equation}
Here, $|\hat{\bbox{p}},j_{3}\rangle\in \Phi^{\times}$ are the
eigenkets of the 4-velocity operator $\hat{P}_{\mu}=P_{\mu}M^{-1}$
and $\phi_{j_{3}}(\hat{\bbox{p}})=\langle j_{3}\hat{\bbox{p}}|\phi\rangle$
represents the 4-velocity distribution of the vector $\phi$.
The 4-velocity eigenvectors are often
more useful for physical reasoning, because 4-velocities seem to
fulfill to rather good approximation ``velocity super-selection
rules'' which the momenta do not~\cite{ref7}.

The relativistic Gamow vectors will therefore be defined, not as
momentum eigenvectors, but as 4-velocity eigenvectors in the direct
product space of UIR spaces for the decay products of the resonances
$R$. We want to obtain the relativistic Gamow vectors from the pole
term of the relativistic S-matrix in complete analogy to the way the
non-relativistic Gamow vectors were obtained~\cite{Bohm1}. In the
absence of a vector space description of a resonance, we shall also in
the relativistic theory define the unstable particle by the pole of
the analytically continued partial S-matrix with angular momentum
$j=j_R$ at the value
${\mathsf{s}}={\mathsf{s}}_R\equiv(M_R-i\Gamma/2)^2$ of the invariant
mass square variable (Mandelstam variable)
${\mathsf{s}}=(p_1+p_2+\cdots)^2=E_R^2-{\bbox p}_R^2$, where $p_1$,
$p_2$,\ldots are the momenta of the decay products of
$R$~\cite{Eden}. This means that the mass $M_R$ and lifetime
$\hbar/\Gamma_R$ or the complex invariant mass
$w_{R}=(M_{R}-i\Gamma/2)=\sqrt{{\mathsf{s}}_{R}}$, in addition to spin
$j_R$, are the intrinsic properties that define a quasistable
relativistic particle~\footnote{Conventionally and equivalently one
often writes
\[
{\mathsf{s}}_R\equiv
M_\rho^2-iM_\rho\Gamma_\rho=
M_R^2\left(1-\frac{1}{4}
\left(\frac{\Gamma_R}{M_R}\right)^2\right)-iM_R\Gamma_R
\]
and calls
$M_\rho=M_R\sqrt{1-\frac{1}{4}\left(\frac{\Gamma_R}{M_R}\right)^2}$
the resonance mass and $\Gamma_\rho=\Gamma_R\left(1-\frac{1}{4}
\left(\frac{\Gamma_R}{M_R}\right)^2\right)^{-1/2}$ its width. For the
$\rho$ meson $\left(\frac{\Gamma_R}{M_R}\right)^2\approx0.03$ and for
most other resonances it is an order of magnitude or more smaller than
this.}.

In order to make the analytic continuation of the partial S-matrix
with angular momentum $j$, we need the angular momentum basis vectors
\begin{eqnarray}
&|\hat{\bbox p}j_3(wj)\rangle=
\int\frac{d^3\hat{p}_1}{2\hat{E}_1}
\frac{d^3\hat{p}_2}{2\hat{E}_2}
|\hat{\bbox p}_1\hat{\bbox p}_2[m_1m_2]\rangle
\langle\hat{\bbox p}_1\hat{\bbox p}_2[m_1m_2]|\hat{\bbox p}j_3(wj)\rangle
\label{eq11}\\
&\mbox{for any $(m_1+m_2)^2\leq w^2<\infty$ \,\,\,\,$j=0,1,\ldots$}
\nonumber
\end{eqnarray}
in the direct product space of the decay products of the resonance $R$
\begin{equation}
{\mathcal H}\equiv
{\mathcal H}(m_1,0)\otimes
{\mathcal H}(m_2,0)=
\int_{(m_1+m_2)^2}^{\infty}
d{\mathsf{s}}
\sum_{j=0}^{\infty}\oplus{\mathcal H}({\mathsf{s}},j),
\label{eq12}
\end{equation}
where ${\mathsf{s}}=w^{2}$, the Mandelstam variable defined above.

For simplicity, we have assumed here that there are two decay products,
$R\rightarrow\pi_1+\pi_2$ with spin zero, described by the irreducible
representation spaces ${\mathcal H}^{\pi_i}(m_i,s_i=0)$~\footnote{Though
our discussions apply with obvious modifications to the general case of
\[
1+2+3+\cdots\rightarrow R_i\rightarrow
1^\prime+
2^\prime+
3^\prime+\cdots,
\]
these generalizations lead to enormously more complicated equations.}
of
the Poincar\'e group ${\mathcal P}$.

The kets $|\hat{\bbox p}j_3(wj)\rangle$ are eigenvectors of the 4-velocity
operators
\begin{equation}
\hat{P}_\mu=(P^1_\mu+P^2_\mu)M^{-1},\,\,\,\,
M^2=(P^1_\mu+P^2_\mu)({P^1}^\mu+{P^2}^\mu)
\label{eq13}
\end{equation}
with eigenvalues
\begin{equation}
\hat{p}^\mu=\left(
\begin{array}{c}
\hat{E}=\frac{p^0}{w}=\sqrt{1+\hat{\bbox p}^2}\\
\hat{\bbox p}=\frac{\bbox p}{w}
\end{array}
\right)
\,\,\,\,\,\,\mbox{and}\,\,\,\,
w^2={\mathsf{s}}.
\end{equation}
In here $\hat{P}^i_\mu$ are the 4-velocity operators in the one
particle spaces
${\mathcal H}^{\pi_i}(m_i,s_i)$ with eigenvalues
$\hat{p}^i_\mu=\frac{p^i_\mu}{m_i}$.

To obtain the Clebsch-Gordan coefficients $\langle \hat{\bbox
p}_{1}\hat{\bbox p}_{2}[m_1,m_2]|\hat{\bbox p}j_{3}(wj)\rangle$
in~(\ref{eq11}), one follows the same procedure as given in the
classic papers~\cite{Wight,Joos,Macf} for the Clebsch-Gordan
coefficients for the Wigner (momentum) basis. This has been done
in~\cite{Ref12}. The result is
\begin{eqnarray}
&\langle\hat{\bbox p}_1\hat{\bbox p}_2[m_1,m_2]|\hat{\bbox p}j_3(wj)\rangle=
2\hat{E}(\hat{\bbox p})\delta^3(\bbox{p}-\bbox{r})\delta(w-\epsilon)
Y_{jj_3}({\bbox e})\mu_j(w^2,m_1^2,m_2^2)
\label{eq15}\\
&\mbox{with}\,\,\,\epsilon^2=r^2=(p_1+p_2)^2,
\,\,\,r=p_1+p_2,\nonumber
\end{eqnarray}
where $\mu_j(w^2,m_1^2,m_2^2)$ is a function that fixes the
$\delta$-function ``normalization'' of $|\hat{\bbox
p}j_3(wj)\rangle$. The unit vector ${\bbox e}$ in~(\ref{eq15}) is
chosen to be in the c.m.\ frame the direction of $\hat{\bbox
p}_1^{\textrm{\scriptsize cm}} =-\frac{m_2}{m_1}\hat{\bbox
p}_2^{\textrm{\scriptsize cm}}$.

The normalization of the basis vectors~(\ref{eq11}) is chosen to be
\begin{eqnarray}
&\langle\hat{\bbox p}^\prime j^\prime_3(w^\prime j^\prime)
|\hat{\bbox p}j_3(wj)\rangle=
2\hat{E}(\hat{\bbox p})\delta(\hat{\bbox p}^\prime-\hat{\bbox p})
\delta_{j_3^\prime j_3}\delta_{j^\prime j}
\delta({\mathsf{s}}-{\mathsf{s}}^\prime)
\label{eq16}\\
&\mbox{where}\,\,\,\,
\hat{E}(\hat{\bbox p})=\sqrt{1+\hat{\bbox p}^2}=\frac{1}{w}
\sqrt{w^2+\bbox{p}^2}\equiv
\frac{1}{w}E({\bbox p},w).\nonumber
\end{eqnarray}
This determines the weight function $\mu_j(w^2,m_1^2,m_2^2)$ to
\begin{equation}
\left| \mu_j(w^2,m_1^2,m_2^2)\right|^{2}=
\frac{2m_1^2m_2^2w^2}{\sqrt{\lambda(1,(\frac{m_1}{w})^2,(\frac{m_2}{w})^2)}}
\label{eq17}
\end{equation}
where $\lambda$ is defined by~\cite{Wight}
\begin{equation}
\lambda(a,b,c)=a^2+b^2+c^2-2(ab+bc+ac).
\label{eq18}
\end{equation}
The basis vectors~(\ref{eq11}) are the eigenvectors of the free
Hamiltonian $H_0=P^1_0+P^2_0$
\begin{equation}
H_0^{\times}|\hat{\bbox p}j_3(wj)\rangle=
E|\hat{\bbox p}j_3(wj)\rangle,\,\,\,\,\,\,\,
E=w\sqrt{1+\hat{\bbox{p}}^2}.
\label{eq19}
\end{equation}
The Dirac-Lippmann-Schwinger scattering states are obtained, in
analogy to~(\ref{eq6}) (cf. also~\cite{Weinberg} sect.~3.1) by:
\begin{equation}
|\hat{\bbox p}j_3(wj)^\pm\rangle=
\Omega^\pm|\hat{\bbox p}j_3(wj)\rangle
\label{eq20}
\end{equation}
where $\Omega^\pm$ are the M{\o}ller operators. For the basis vectors
at rest, (\ref{eq20}) is given by the solution of the
Lippmann-Schwinger equation
\begin{equation}
|\bbox{0}j_3(wj)^\pm\rangle=
\left(1+
\frac{1}{w-H\pm i\epsilon}V
\right)
|\bbox{0}j_3(wj)\rangle.
\label{eq21}
\end{equation}
They are eigenvectors of the exact Hamiltonian $H=H_0+V$
\begin{equation}
H|\bbox{0}j_3(wj)^\pm\rangle=
\sqrt{{\mathsf{s}}}|\bbox{0}j_3(wj)^\pm\rangle,
\,\,\,\,
(m_1+m_2)^2\leq {\mathsf{s}}<\infty.
\label{eq22}
\end{equation}
The vectors $|\hat{\bbox p}j_3(wj)^\pm\rangle$ are obtained from the basis
vectors at rest $|\bbox{0}j_3(wj)^\pm\rangle$ by the boost
(rotation-free Lorentz transformation) ${\mathcal U}(L(\hat{p}))$
whose parameters are the 4-velocities $\hat{p}^{\mu}$. The generators 
of the Lorentz transformations are the
interaction-incorporating observables
\begin{equation}
P_0=H,
\,\,\,\,
P^m,
\,\,\,\,
J_{\mu\nu},
\label{eq23}
\end{equation}
i.e., the exact generators of the Poincar\'e group (\cite{Weinberg}
sec.\ 3.3). These vectors $|\hat{\bbox p}j_{3}(wj)^{\pm}\rangle$
in (\ref{eq20}), or $|\bbox{0}j_{3}(wj)^{\pm}\rangle$ in (\ref{eq21})
when boosted by ${\mathcal U}(L(\hat{p}))$, 
which for the fixed value
$[jw]$ span an irreducible representation space of the Poincar\'e
group with the ``exact generators'', will be used for the definition
of the relativistic Gamow vectors.

The relativistic Gamow kets are obtained by analytically continuing
the Dirac-Lippmann-Schwinger kets~(\ref{eq21}) or~(\ref{eq20}) into
the second (or higher) sheet of the $j_R$-th partial wave S-matrix
$S_{j_R}({\mathsf{s}})$ to the position of the resonance pole at
${\mathsf{s}}_R=(M_R-i\Gamma/2)^2$. This can be done for any value of
$\hat{\bbox p}$ (and $j_3$), e.g., for $\hat{\bbox p}={\bbox 0}$. In
complete analogy to the non-relativistic case one obtains the
relativistic kets
\begin{equation}
|\hat{\bbox p}j_3({\mathsf{s}}_Rj_R)^{-}\rangle=
\frac{i}{2\pi}
\int_{-\infty_{II}}^{+\infty_{II}}d{\mathsf{s}}
|\hat{\bbox p}j_3({\mathsf{s}} j_R)^{-}
\rangle\frac{1}{{\mathsf{s}}-{\mathsf{s}}_R}.
\label{eq24}
\end{equation}
The Lorenz transformations $\Lambda$ are represented by unitary
operators ${\mathcal U}(\Lambda)$:
\begin{equation}
{\mathcal U}(\Lambda)|\hat{\bbox p}j_3({\mathsf{s}}_R j_R)^{-}\rangle=
\sum_{j_3^\prime}D^{j_{R}}_{j_{3}^{\prime}j_{3}}
({\mathcal R}(\Lambda,\hat{p}))
|\bbox{\Lambda}\hat{\bbox{p}}j_3^\prime({\mathsf{s}}_R j_R)^{-}\rangle,
\label{eq25}
\end{equation}
where ${\mathcal R}(\Lambda,\hat{p})=L^{-1}(\Lambda \hat{p})\Lambda
L(\hat{p})$ is the Wigner rotation.  In particular for the rotation
free Lorentz boost
\begin{equation}
{\mathcal U}(L(\hat{p}))
|\hat{\bbox p}={\bbox 0},j_3({\mathsf{s}}_R j_R)^{-}\rangle=
|\hat{\bbox p}j_3({\mathsf{s}}_R j_R)^{-}\rangle
\label{eq26}
\end{equation}
because the boost is a function of the real $\hat{p}^\mu$ and not
of the complex $p^\mu$:
\begin{equation}
L^\mu_{\hphantom{\mu}\nu}=
\left(
\begin{array}{cc}
\frac{p^0}{m}&-\frac{p_n}{m}\\
\frac{p^k}{m}&\delta^k_n-\frac{\frac{p^k}{m}\frac{p_n}{m}}{1+\frac{p^0}{m}}
\end{array}
\right),
\,\,\,\,
L(\hat{p})
\left(
\begin{array}{c}
1\\
0\\
0\\
0
\end{array}
\right)
=\hat{p}.
\label{eq27}
\end{equation}

The relativistic Gamow kets~(\ref{eq24}) are generalized eigenvectors
of the invariant mass squared operator $M^2=P_\mu P^\mu$ with
eigenvalue ${\mathsf{s}}_R$
\begin{equation}
\langle\psi^-|M^2|\hat{\bbox p}j_3({\mathsf{s}}_Rj_R)^{\,\,\,\,-}\rangle=
{\mathsf{s}}_R\langle\psi^-|\hat{\bbox p}j_3
({\mathsf{s}}_Rj_R)^{\,\,\,\,-}\rangle
\quad \text{for every } \psi^-\in\Phi_+\subset
{\mathcal H}\subset\Phi^\times_+.
\label{eq28}
\end{equation}
To prove~(\ref{eq28}) from~(\ref{eq24}), and also to
obtain~(\ref{eq24}) from the pole term of the S-matrix one needs to
use the properties of the Hardy class space $\Phi_+$.  The continuous
linear combinations of the 4-velocity kets (\ref{eq24}) with an
arbitrary 4-velocity distribution function
$\phi_{j_{3}}(\hat{\bbox{p}})\in {\cal S}$ (Schwartz space),
\[
\psi^{\rm G}_{j_{R}{\mathsf{s}}_{R}}=
\sum_{j_{3}}\int \frac{d^{3}\hat{p}}{2\hat{p}^{0}}
|\hat{\bbox{p}}j_{3}({\mathsf{s}}_{R},j_{R})^{-}
\rangle \phi_{j_{R}}(\hat{\bbox{p}}),
\]
also represents relativistic Gamow states with the complex
mass ${\mathsf{s}}_{R}=(M_{R}-i\Gamma/2)^{2}$ and have 
the representation~(\ref{eq24}).

This then implies, much like in the RHS theory of non-relativistic
Gamow vectors, that the time translation of the decaying state is
represented by a semigroup, e.g., at rest,
\begin{equation}
\textrm{e}^{-iH\tau}
|\hat{\bbox p}=\bbox{0},j_3({\mathsf{s}}_R j_R)^{-}\rangle=
\textrm{e}^{-im_R\tau}
\textrm{e}^{-\Gamma\tau/2}
|\hat{\bbox p}=\bbox{0},j_3({\mathsf{s}}_R j_R)^{-}\rangle
\,\,\,\,
\text{for }\tau\geq0\,\,\,\textrm{only}
\label{eq29}
\end{equation}
where $\tau$ is time in the rest system.

Thus relativistic Gamow states are representations of ${\mathcal P}$
with spin $j_R$ and complex mass
${\mathsf{s}}_R=(M_{R}-i\Gamma/2)^2=m_\rho^2-im_\rho\Gamma_\rho$, for
which the Lorentz subgroup is unitarily represented. They are obtained
from the resonance pole of the relativistic partial S-matrix
$S_{j_R}({\mathsf{s}})$, have an exponential time evolution at rest
with lifetime $\hbar/\Gamma$ and have -- due to their association to
the S-matrix pole and to the Hardy class spaces -- a Breit-Wigner
energy distribution
\begin{equation}
a_j=\frac{Bm\Gamma}{{\mathsf{s}}-(m-i\frac{\Gamma}{2})^2},
\,\,\,\,
-\infty_{II}<{\mathsf{s}}<+\infty_{II}.
\label{eq30}
\end{equation}
These are all features which one may welcome or accept for states that
are to describe relativistic resonances. In addition, they have a
semigroup time evolution expressing a fundamental quantum mechanical
arrow of time.

We gratefully acknowledge support of the Welch Foundation and of CoNaCyT
(Mexico).


\begin{thebibliography}{99}
\bibitem{ref5}
The projectively equivalent irreducible representations $(m,E,s)\sim
(m,0,s)\equiv [m,s]$ of the Galilei group are characterized by
non-relativistic mass $m$ and spin $s$. In the direct product of
$[m_1,s_1]\otimes[m_2,s_2]$, the two spins are combined (using $SO(3)$
Clebsch-Gordan coefficients) to give the total spin $s$. Then, using
orbital angular momentum $l$ for the relative orbital motion, one
couples $s$ with $l$ to form the total angular momentum $j$ with the
3-component $j_3$. This gives the basis vectors
$|E^{\textrm{\scriptsize tot}},{\bbox p}jj_3(s,l)\rangle$ 
defined by their scalar
products in the center of mass frame ${\bbox p}={\bbox 0}$ (Clebsch-Gordan
coefficients of the Galilei group) with the basis vector
$|{\bbox p}_1,s_1,(s_1)_3\rangle\otimes
|{\bbox p}_2,s_2,(s_2)_3\rangle=
|{\bbox p}_1,(s_1)_3,{\bbox p}_2,(s_2)_3(s_1,s_2)\rangle$
\begin{eqnarray}
\langle{\bbox p}_1,(s_1)_3,{\bbox p}_2,(s_2)_3(s_1,s_2)|
E^{\textrm{\scriptsize tot}},{\bbox p}jj_3(s,l)\rangle
&=&N\delta^3({\bbox{p}-\bbox{p}_1-\bbox{p}_2})
\delta(E^{\textrm{\tiny tot}}-\frac{p_1^2}{2m_1}
-\frac{p_1^2}{2m_1})\nonumber\\
&&\times\sum_{s_3,l_3}
\langle s_1(s_1)_3s_2(s_2)_3|ss_3\rangle
\langle ll_3ss_3|jj_3\rangle
Y_{ll_3}(\bbox{e})\nonumber
\end{eqnarray}
where $\bbox{e}$ is the unit vector in the direction
of $\bbox{p}$ and $N$ is a suitable normalization factor.
\end{thebibliography}
\end{document}